%% file: direction_reconstruction.tex
\documentclass[a4paper,12pt]{article}
\pdfoutput=1 % arxiv guideline
\usepackage[pdftex]{graphicx}
\usepackage[T2A]{fontenc}
\usepackage[utf8]{inputenc}
\usepackage[english]{babel}
\usepackage{amssymb,amsfonts,amsmath,mathtext}
\usepackage{cite,enumerate}
\usepackage{float,indentfirst}
\usepackage{graphicx}
\usepackage{subfigure}
\usepackage[nottoc]{tocbibind}
\usepackage[a-1b]{pdfx}
\usepackage{inputenc}
\hypersetup{pdfencoding=unicode}
\inputencoding{utf8}
\makeatother
\hypersetup{pdfa,bookmarks,breaklinks=true}
\usepackage{bookmark}
\usepackage{esint} % for \oint

\newcommand{\Rd}{\mathbb R^\mathrm d}
\newcommand{\R}[1]{\mathbb R^\mathrm #1}
\newcommand{\ud}{\mathrm{d}}
\renewcommand{\Im}{\mathop{\mathrm{Im}}\nolimits} % from Lvovsky
\renewcommand{\Re}{\mathop{\mathrm{Re}}\nolimits}

\begin{document}
\input{titlepage}

\clearpage

% add a pdf bookmark for Contents.
\hypertarget{tocpage}{}
\tableofcontents
\bookmark[dest=tocpage,level=1]{Contents}

\setcounter{secnumdepth}{-1}
\include{introduction}

\setcounter{secnumdepth}{2}
\include{exp}

\include{gauss}

\setcounter{secnumdepth}{-1}
\section{Conclusions}

In this work a new 
approach to solving 
the problem in~\cite{Chooz99} was proposed.
The precision of the sample mean estimator was calculated
analytically for the offset exponential and normal distributions 
both for a finite sample and for limiting cases.

Even though the original applied problem concerned the exponential distribution,
the normal distribution 
was found to be
also useful 
because of the central limit theorem~\cite{CLT}.
It was shown explicitly how the distribution of the sample mean 
of the exponential pdf converges near the mode to the normal distribution.

While the normal distribution is tractable easier and has simpler formulae 
for the distribution of the sample mean and for the directional CDF,
the exponential distribution has richer mathematical properties.
While the distribution of the convolution of normal pdfs depends only on one
combination of parameters, for the exponential distribution this is not the case.
While the normal distribution is stable, the exponential one is not.
Geometric techniques were used to 
deal with
the limiting case of the 
exponential distribution. 
It was shown that 
the spherical projection of the sample mean 
of the exponential distribution
has connections with hypergeometric functions and modified Bessel functions.

In this study we didn't concern other estimators, 
such as MLEs or the mean of the sample's projection on the sphere. 
Note that in~\cite{Chooz99} it was stated that the mean of unit vectors 
is a more precise estimator than the arithmetic sample mean. 
It might also be useful for mathematical applications to study 
the normal and exponential distributions in dimensions other than three.

\subsection*{Acknowledgements}
I would like to express my gratitude to the Independent University of Moscow,
which largely formed my mathematical way of thinking and 
taught me quite a broad area in advanced mathematics 
(even though this thesis has least to do with what I was taught at IUM).
The years that
I 
remember 
most at IUM
were the first ones, 
when we solved lots of problems and discussed them personally
with mathematicians.

I'm very thankful to professor Alexander Mikhailovich Chebotarev, 
who gave a course on statistics at my primary university and
who agreed to be my scientific adviser at IUM, 
read this thesis, provided useful suggestions and independently checked the results 
\ref{E_CDF_cos_theta_final} and \ref{CDF_G_exact}.

I'm also thankful to all my mathematics 
professors
at my primary university, 
which was called the Institute of Natural Sciences and Ecology 
and now is the 10th faculty of the Moscow Institute of Physics and Technology, 
and which gave me very good fundamental and more 
advanced knowledge of mathematics;
and at my physical and mathematical school 1189 
and earlier. 

I would like to thank my current employer,
the Institute for Nuclear Research of the Russian Academy of Sciences,
my scientific adviser there Valery Vitalievich Sinev,
and my mother Arsenia Nikitenko.

\end{document}

%% file: titlepage.tex
\begin{titlepage}

% Titlepage template adapted from 
% http://ctan.org/pkg/titlepages
% by Peter R. Wilson.
\newlength{\drop}
\newcommand*{\titlePW}{\begingroup% PW Thesis
\ttfamily
\drop=0.1\textheight

\centering
{\Large%\bfseries
ESTIMATION OF THE DIRECTIONAL PARAMETER
OF THE OFFSET EXPONENTIAL AND NORMAL DISTRIBUTIONS
IN THREE-DIMENSIONAL SPACE
USING THE SAMPLE MEAN%
\par}

\begin{center}
\begin{figure*}[h]
\includegraphics[width=\textwidth]{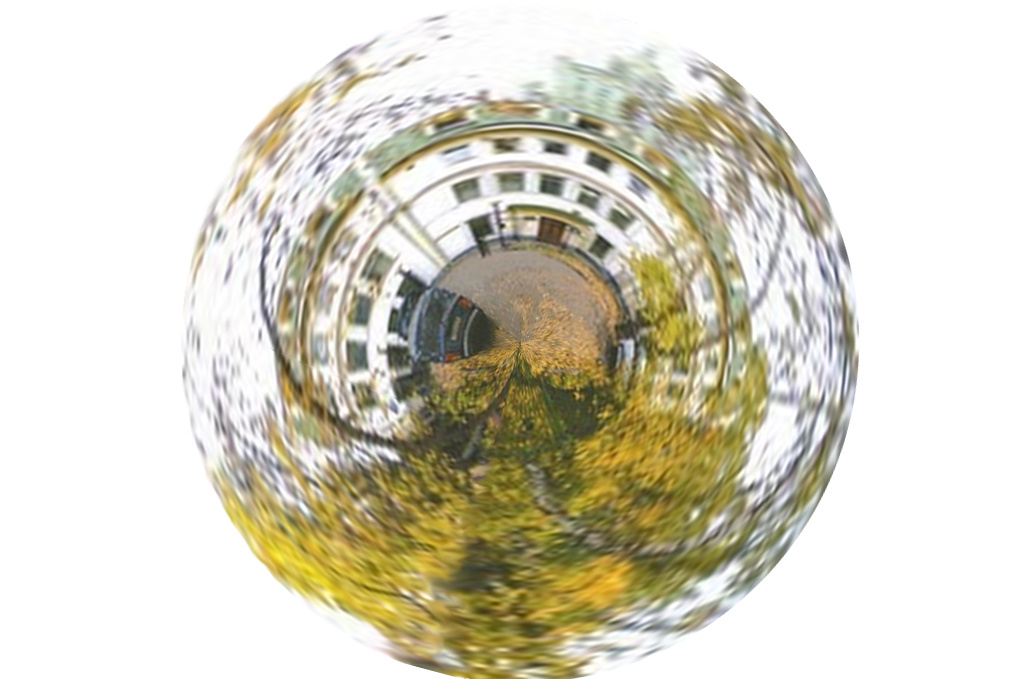}
\end{figure*}
\end{center}

{\Large 
YAROSLAV NIKITENKO
\par}
\vspace*{0.3 \baselineskip}
\raggedright
{\Large
\center{
Thesis submitted at the 
\\
Independent University of Moscow%
}
{
\par
\vspace*{0.3 \baselineskip}
\flushleft{ Scientific adviser: %\\
\vspace*{-0.3 \baselineskip}
\center{Prof. Alexander Mikhailovich Chebotarev\par}}
\vspace{0.7cm}
\center{2015\par}
}
}%end Large
\vfill
\null
\endgroup}

\titlePW

% not sure whether these two variables are needed for arXiv processing.
\title{
Estimation of the directional parameter
\\
of the offset exponential and normal distributions
\\
in three-dimensional space
\\
using the sample mean
}

\author{Yaroslav Nikitenko
  \thanks{\href{mailto:metst13@gmail.com}{metst13@gmail.com}}
}

%\maketitle

\end{titlepage}

%% file: introduction.tex
\section{Introduction}
The problem originated from neutrino physics \cite{Chooz99}.

We consider a set of vectors in 3-dimensional Euclidean space $\mathbb R^3$.

We make a parametric assumption that this set is a sample 
of independent identically distributed variables, 
where a parameter of the distribution is
%from a known distribution with the common parameter which is 
a direction. 

As our estimator we take the direction of the arithmetic mean of the sample.
This allows a simpler mathematical treatment compared to other possible estimators,
since the sum of variables corresponds to the convolution of their pdfs, 
and this can be calculated in a standard way using the Fourier transform.

Our goal is to find the distribution of the estimate in order to calculate confidence sets on the sphere, which we consider the precision of the estimator. 
We study both the exact case for finite samples and asymptotic cases, e.g. for number of events large.

Our parametric models are the exponential distribution, section \ref{exp_distribution}, and the normal (Gaussian) distribution, section \ref{normal_distribution}.

These results are new compared to previous studies. The author was unable to find directional results for the exponential distribution. 
In physical articles only the limiting case of large number of events is usually considered \cite{Chooz99}.
Mathematical literature on directional statistics usually deals with distributions on spheres \cite{MardiaJupp}, while in our case we have complete 3-dimensional information.

This work was written for those who are not necessarily statisticians or mathematicians
but have met the problems treated here.
%in their applications
Therefore the author attempted to use only the basic facts 
from mathematical undergraduate courses 
and introduced in detail more advanced notions when they were used. 
All the references used in this work can be found on the internet.

\subsection{Convolution of pdfs using the Fourier transform}

The probability density function (pdf)
$f(\mathbf r)$
%$f_{\mathbf r_1+\mathbf r_2}(\mathbf r)$
of the sum of two independent variables 
$\mathbf r_1+\mathbf r_2$ 
in $\Rd$
is given by the \textit{convolution} of their pdfs:

$$
f_{\mathbf r_1+\mathbf r_2}(\mathbf r) = 
(f_1 * f_2)(\mathbf r) = 
\int_{\Rd} f_1(\mathbf r') f_2(\mathbf r - \mathbf r')
\,\ud \mathbf r'
$$

We denote the \textit{Fourier transform}
\footnote{this is similar to the \textit{characteristic function} in probability theory, 
the latter is complex conjugate and without the factor $(2\pi)^{\frac d2}$
} 
of a function $f(\mathbf r)$ as 

$$
\hat f(\mathbf p) = 
\int_{\Rd} 
\frac{e^{-i\mathbf{pr}}}{(2\pi)^{d/2}}
f(\mathbf{r})
\,\ud^d \mathbf r,
\label{Fourier}
$$

and the inverse Fourier transform of $f$ as $\tilde f$ 
(therefore $\tilde{\hat f}(x) \equiv f(x)$). 
With this definition the inverse Fourier transform operator is complex conjugate 
to the direct Fourier transform operator.
Then

\begin{align}
\widehat{f * g}(\mathbf p)
&=
\int_{\Rd} 
\mathrm{d}^d \mathbf r
\int_{\Rd} 
    \frac{e^{-i\mathbf{pr}}}{(2\pi)^{d/2}}
    f(\mathbf r') g(\mathbf r - \mathbf r')
\,\mathrm{d}^d \mathbf r'
\nonumber
\\
&=
\int_{\Rd} 
\frac{e^{-i\mathbf{pr}'}}{(2\pi)^{d/2}}
f(\mathbf r')
\,\mathrm{d}^d \mathbf r'
\int_{\Rd} 
e^{-i\mathbf p(\mathbf r - \mathbf r')} g(\mathbf r - \mathbf r')
\,\mathrm{d}^d \mathbf r
\nonumber
\\
&=
(2\pi)^{d/2}\hat f(\mathbf p) \hat g(\mathbf p).
\end{align}

This is a well-known property of the Fourier transform, 
that it maps the \textit{convolution} of two pdfs to the \textit{product} of their Fourier transforms.

Therefore the Fourier transform of the convolution 
of $n$ distributions $f$ is

\begin{equation}
\label{fenFourierGen}
\hat f_n(\mathbf p) 
=
(\hat f (\mathbf p))^n (2\pi)^{\frac {(n-1)d}2}
\end{equation}

%% file: exp.tex
\section{Exponential distribution}
\label{exp_distribution}
\subsection{Introduction}

Exponential distribution appeared in the author's studies connected to the problem in~\cite{Chooz99}.
The observed pdf at large $x$ deviations was similar to $\sim e^{-\frac xl}$. 
To be spherically symmetric the pdf should be proportional to 
$\sim e^{-\frac rl}$.
To calculate the normalisation factor, we take the integral

\begin{equation*}
\int_{-\infty}^{\infty}\int_{-\infty}^{\infty}\int_{-\infty}^{\infty}
e^{-\frac{\sqrt{x^2 + y^2 + z^2}}{l}}
\,\ud x\,\ud y\,\ud z = 
4\pi \int_{0}^{\infty} r^2 e^{-\frac{r}{l}}
\,\ud r 
=
8 \pi l^3
\end{equation*}

Hence the offset exponential probability density function in 3-dimensional space is

\begin{equation}
f_e(x,y,z|x_0, y_0, z_0)
= \frac 1{8 \pi l^3} 
e^{-\frac{\sqrt{(x-x_0)^2 + (y-y_0)^2 + (z-z_0)^2}}l}
\label{f_e}
\end{equation}

\subsubsection
[Fourier transform and convolutions]
{Fourier transform of $f_e$ and its convolutions}

In order to calculate the Fourier transform of $f_e$, we calculate the integral

\begin{eqnarray*}
\iiint_{\R3}
e^{-i\mathbf{pr}}e^{-\frac{\sqrt{(\mathbf r - \mathbf r_0)^2}}{l}}
\, \ud^3\mathbf r
& = &
\iiint_{\R3}
e^{-i\mathbf{pr}_0}
e^{-i\mathbf{pr}'}
e^{-\frac {r'}l}
\, \ud^3\mathbf r' \\
& = &
2\pi e^{-i\mathbf{pr}_0}
\int_0^\infty 
\int_0^\pi 
e^{-ipr'\cos\theta - \frac{r'}l} 
\sin \theta \,\ud \theta
\, r'^2
\, \ud r',
\end{eqnarray*}

the inner integral on $\theta$

\begin{equation}
\int_{-1}^{1} e^{-ipr'\cos\theta}\,\ud\cos\theta
= \frac 1{-ipr'}(e^{-ipr'} - e^{ipr'})
= \frac 1{ipr'}(e^{ipr'} - e^{-ipr'}),
\label{exptheta}
\end{equation}

and the outer integral on $r'$ with one of the complex conjugate exponents

\begin{displaymath}
\int_0^\infty 
r' e^{ipr' - \frac{r'}{l}}
\,\ud r'
=\left(r' = \frac{r}{\frac 1l - ip}\right)
=
\frac 1{(\frac 1l - ip)^2}
\int_0^\infty 
r e^{-r}\,\ud r
=
\frac{1}{(\frac{1}{l} - ip)^2};
\end{displaymath}

combining the two conjugate integrals, we obtain 

\begin{displaymath}
\frac 1{ip}
( \frac{1}{(\frac{1}{l} - ip)^2} - c.c.) 
=
\frac{(\frac 1l+ip)^2 - (\frac 1l - ip)^2}
 {ip(\frac 1l - ip)^2(\frac 1l + ip)^2}
=
\frac{\frac{4ip}l}
 {ip(\frac 1{l^2} + p^2)^2}
=
\frac 4{l(\frac 1{l^2} + p^2)^2},
\end{displaymath}

therefore, taking into account the normalisation factor 
$\frac 1{8\pi l^2}$
and the factor $(2\pi)^{-\frac 32}$ from the Fourier transform, 

\begin{equation}
\hat{f}_e(\mathbf p) = 
\frac 1{8\pi l^3}\frac{8\pi}
 {(2\pi)^\frac 32 l (\frac 1{l^2} + p^2)^2}
e^{-i\mathbf{pr}_0}
=
\label{feFourier}
\frac{e^{-i\mathbf{pr}_0}}
 {(2\pi)^\frac 32(1 + l^2p^2)^2}
\end{equation}

From~\ref{fenFourierGen} and~\ref{feFourier} we can learn the Fourier transform 
of the convolution of $n$ exponential distributions:

\begin{equation}
\hat f_n(\mathbf p) 
= 
\frac{e^{-i\mathbf{pr}_0n}}
{(2\pi)^{\frac 32}(1 + l^2p^2)^{2n}}
.
\end{equation}

Thereby the convolution of $n$ exponential distributions is

\begin{equation*}
f_n(\mathbf r) 
=
\tilde{\hat{f}}_n(\mathbf p)
=
\iiint_{\R3} 
\frac{e^{i\mathbf{pr}}}
 {(2\pi)^\frac 32}
\hat f_n(\mathbf p)
\,\ud^3\mathbf p 
=
\frac 1{(2\pi)^3}
\iiint_{\R3} 
\frac{e^{i\mathbf{p}(\mathbf r - \mathbf r_0 n)}}
 {(1 + l^2 p^2)^{2n}}
\,\ud^3\mathbf p 
,
\end{equation*}

choosing spherical coordinates with the $z$ axis along $\mathbf r - n \mathbf r_0$,
the exponent becomes $e^{i p |\mathbf r - n \mathbf r_0| \cos \theta}$, 
and
using~\ref{exptheta} %(mind the complex conjugation)
,

\begin{eqnarray*}
f_n(\mathbf r) 
& = &
\frac 1{(2\pi)^2}
\int_0^\infty
\frac{e^{ipr|\mathbf r - n \mathbf r_0|} - e^{- ipr|\mathbf r - n \mathbf r_0|}}
 {ip|\mathbf r - n \mathbf r_0|}
\frac 1{(1 + l^2 p^2)^{2n}}
p^2 \,\ud p 
=
\\
& = &
\frac 1{2 \pi^2 |\mathbf r - n \mathbf r_0|}
\int_0^\infty
\frac{p \sin (p|\mathbf r - n \mathbf r_0|)}
 {(1 + l^2 p^2)^{2n}}
\,\ud p 
.
\end{eqnarray*}

Substituting inside the integral $p = \frac xl$,

\begin{equation}
f_n(\mathbf r) 
=
\frac 1{2 \pi^2 l^2 |\mathbf r - n \mathbf r_0|}
\int_0^\infty
\frac{x \sin (x \frac {|\mathbf r - n \mathbf r_0|}l) }
 {(1 + x^2)^{2n}}
\,\ud x 
.
\end{equation}

\subsubsection
{Distribution of the sample mean \texorpdfstring{$E_n$}{Eₙ}}

A statistic useful in practical applications is 
$\mathbf r_n = \frac{\mathbf r}n$
, the arithmetic mean of $\mathbf r$.
We can calculate the probability density function $E_n(\mathbf r_n)$ 
of the random variable $\mathbf r_n$ 
and, using the conservation of probability under the change of variables 
$E_n(\mathbf r_n)\,\ud^3\mathbf r_n = f_n(\mathbf r)\,\ud^3\mathbf r$
, we obtain 
$ E_n(\mathbf r_n) = n^3 f_n(n\mathbf r_n) $

\begin{equation}
E_n(\mathbf r_n) 
=
\frac {n^2}{2 \pi^2 l^2 |\mathbf r_n - \mathbf r_0|}
\int_0^\infty
\frac{x \sin (x \frac {n |\mathbf r_n - \mathbf r_0|}l) }
 {(1 + x^2)^{2n}}
\,\ud x. 
\label{E_n}
\end{equation}

The integral can be calculated analytically using the formula 
3.737(2) from~\cite{GradRyzh}
$[a > 0, \Re \beta > 0 ]$:

\begin{equation}
\int_0^\infty 
\frac{x\sin (ax)\,\ud x}
     {(x^2 + \beta^2)^{n+1}}
=
\left\{
\begin{aligned} &
\frac{\pi a e^{-a\beta}}
 {2^{2n}n!\beta^{2n-1}}
\sum_{k=0}^{n-1} 
\frac{(2n-k-2)!(2a\beta)^k}
     {k!(n-k-1)!}
\label{GR_int} 
\\ & 
\frac \pi2 e^{-a\beta}
\qquad \qquad \qquad \quad 
\left[n=0, \beta \geq 0\right]
\end{aligned}
\right.
\end{equation}

Combining \ref{E_n} and \ref{GR_int} 
, we obtain

\begin{equation}
E_n(\mathbf r_n) 
=
\frac {n^3}{\pi l^3}
\frac{e^{-\frac nl |\mathbf r_n - \mathbf r_0|}}
     {2^{4n-1}(2n-1)!}
\sum_{k=0}^{2n-2} 
\frac{(4n-4-k)!(2 \frac nl |\mathbf r_n - \mathbf r_0|)^k}
     {k!(2n-2-k)!}
\label{E_n_full}
\end{equation}

In the case of $n=1$ the sum in~\ref{E_n_full} 
is equal to
$1$ and we obtain~\ref{f_e}.

\subsection
{Properties of \texorpdfstring{$E_n$}{Eₙ}}
In this subsection we study representations of $E_n$ other than~\ref{E_n_full}
and its connection with hypergeometric functions.

\input{exp_integral_GR.tex}

\subsubsection
%[Proof that %the pdf %of the sample mean is properly normalised]
{Proof that \texorpdfstring{$E_n$}{Eₙ} is properly normalised}

The integral $\int_{\R3} E_n(\mathbf r_n)\,\ud^3 \mathbf r_n$ is equal to~$1$,
since $E_n$ is a properly normalised pdf.
Here we calculate it explicitly using the formula~\ref{E_n_full}.

After the parallel shift of $\mathbf{r}_n$ to $\mathbf{r}_0$, 
which doesn't affect the total integral, 
after changing to spherical coordinates and having integrated on the polar angles, we obtain the equality to be proved

\begin{equation*}
\frac{n^3}{\pi l^3} \,
4\pi
\int_0^\infty r_n^2\,\ud r_n
\frac{e^{-\frac nl r_n}}
     {2^{4n-1}(2n-1)!}
\sum_{k=0}^{2n-2} 
\frac{(4n-4-k)!(2 \frac nl r_n)^k}
     {k!(2n-2-k)!}
= 1.
\end{equation*}

Using the integral
$ \int_0^\infty x^n e^{-x} \,\ud x = \Gamma(n+1) = n!$, 
this transforms to

\begin{equation}
\frac 1{2^{4n-3}(2n-1)!}
\sum_{k=0}^{2n-2} 
\frac{(4n-4-k)!2^k(k+2)!}
     {k!(2n-2-k)!}
= 1.
%\equiv 1.
\label{normfact}
\end{equation}

This equality holds for $n=1$. For $n=2$ the left-hand side

$$
\frac 1{2^53!}\left(\frac{4!\,2!}{2!} 
+ \frac{3! \, 2 \!\cdot\! 3!}{1!}
+ \frac{2!\,2^2 \!\cdot\! 4!}{2!}\right) 
= \frac 1{2^5}(4 + 12 + 16) 
= 1.
$$

In the remaining of this subsubsection we prove the identity \ref{normfact}.
Different techniques for calculation of closed forms 
of summations involving binomial coefficients 
can be found in~\cite{GKP}. 
Here we use the method of hypergeometric functions.

The Gaussian (or ordinary) hypergeometric function $_2F_1(a,b;c;z)$ 
is a special function represented by the hypergeometric series 
(\!\!\cite{PBM}, 7.2.1(1))

\begin{eqnarray*}
_2F_1(a,b;c;z) & = & 1 + \frac{ab}c z + 
\frac{a(a+1)b(b+1)}{c(c+1)}\frac{z^2}{2!} + \\
& & + \frac{a(a+1)(a+2)b(b+1)(b+2)}{c(c+1)(c+2)}\frac{z^3}{3!} + \ldots 
\end{eqnarray*}

This can be rewritten using the rising factorial or Pochhammer symbol

\begin{equation}
\begin{split}
(a)_0 & = 1, 
\\
(a)_n & = a(a+1)\ldots(a+n-1),
\label{Pochhammer}
\end{split}
\end{equation}

then

\begin{equation}
_2F_1(a,b;c;z) = \sum_{k=0}^\infty \frac{(a)_k(b)_k}{(c)_k} \frac{z^k}{k!}.
\end{equation}

In case when $a$ or $b$ is a negative integer, only a finite number of terms is non zero. 
Using Pochhammer symbol~\ref{Pochhammer}, we can express

\begin{eqnarray}
\frac{n!}{(n-k)!} = n (n-1) \ldots (n-k+1) = (-1)^k(-n)_k,
\nonumber
\\
(n-k)! = \frac{n!}{(-1)^k(-n)_k}
\label{(n-k)!}
\\
(k+2)! = 1 \cdot 2 \cdot 3 \ldots (k+2) = 2\cdot(3)_k,
\label{(k+2)!}
\end{eqnarray}

The sum in \ref{normfact} can be rewritten as

\begin{eqnarray}
& &
\sum_{k=0}^{2n-2}
\frac{(4n-4-k)!(k+2)!} {(2n-2-k)!}
\frac {2^k}{k!}
\stackrel{\left( \textrm{ \ref{(n-k)!},\ref{(k+2)!} } \right)}{=}
2 \frac{(4n-4)!}{(2n-2)!}
\sum_{k=0}^{2n-2}
\frac{(-2n+2)_k (3)_k}{(-4n+4)_k}
\frac{2^k}{k!}
= 
\nonumber
\\
& & 
\hspace{4.5cm}
=
2 \frac{(4n-4)!}{(2n-2)!}
\,
{}_2F_1(-2n+2,3;-4n+4;2)
\label{2F1(2)}
\end{eqnarray}

This hypergeometric function can be calculated 
using the formula 7.3.8(6) in~\cite{PBM}:

\begin{equation}
_2F_1(-n,a;-2n;2) = 2^{2n} \frac{n!}{(2n)!}
\left(\frac{a+1}2\right)_{\!\!n}
\label{2F1(2)PBM}.
\end{equation}

Substituting \ref{2F1(2)PBM} into \ref{2F1(2)}, we obtain the original equality~\ref{normfact}. %$\blacksquare$

\input{exp_cdf.tex}

\input{exp_cdf_approx}

%% file: exp_integral_GR.tex
\subsubsection
{Calculation of the integral \ref{GR_int}}
\label{exp_integral_GR}

The integral \ref{GR_int} for $\alpha,\beta>0$
can be easily reduced to that with $\beta = 1$:

\begin{equation}
\int_0^\infty 
    \frac{x\sin(\alpha x)}{(x^2 + \beta^2)^{n+1}} 
\ud x
\stackrel{x=\beta y}{=}
\frac 1{\beta^{2n}}
\int_0^\infty 
    \frac{y\sin(\alpha \beta y)}{(y^2 + 1)^{n+1}} 
\ud y
\label{GR_int_1}
\end{equation}

\paragraph{\fbox{$n > 0$}}

The integral \ref{GR_int_1} using integration by parts can be transformed to

\begin{align}
\int_0^\infty 
    \frac{x\sin(a x)}{(x^2 + 1)^{n+1}} 
\ud x
& = 
- \left. \frac 1{2n} \sin(ax) \frac 1{(1+x^2)^n}\right|_0^\infty
+
\frac a{2n}
\int_0^\infty 
    \frac{\cos(a x)}{(x^2 + 1)^{n}}
\ud x
=
\notag
\\
& = 
\frac a{2n}
\int_0^\infty 
    \frac{\cos(a x)}{(x^2 + 1)^{n}}
\ud x
.
\label{GR_cos_int}
\end{align}

To compute this integral we apply complex analysis. We express

\begin{equation}
\int_0^\infty \frac{\cos(ax)}{(x^2+1)^n} \ud x = 
\frac 12 \Re \int_{-\infty}^\infty \frac{e^{iax}}{(x^2+1)^n} \ud x, 
\end{equation}

the integrated function 
is meromorphic on whole complex plane and has poles at $\pm i$.
We close the integration contour in the upper half-plane near infinity,
where the integral is zero, 
and deform the contour into a small circle 
$\mathcal C$ of radius $r$ around the pole $+i$.
We parameterise
$z = i + re^{i\phi}$, then

\begin{equation}
\oint_{\mathcal C} 
    \frac{e^{iaz}}{(z^2+1)^n} 
\ud z
= 
\int_0^{2\pi}
\frac {e^{-a + iar e^{i\phi}} ir e^{i\phi} \ud\phi}
      {(2ire^{i\phi} + r^2e^{2i\phi})^n} 
=
\frac {e^{-a}}{2^n(ir)^{n-1}}
\int_0^{2\pi}
\frac {e^{iar e^{i\phi}} e^{- i(n-1)\phi} \ud\phi}
      {(1 + \frac{r}{2i} e^{i\phi})^n} 
\end{equation}

In order to compute that integral, we decompose 
the integrand
with respect to power series of $e^{i\phi}$
and use the identity

\begin{equation}
\int_0^{2\pi} e^{in\phi} \ud \phi = 
\begin{cases}
0 & \text{if } n \neq 0, \\
2\pi & \text{if } n = 0.
\end{cases}
\end{equation}

Since $r$ can be made sufficiently small, we can decompose the denominator 
using the binomial formula:

\begin{gather}
(1+x)^{-n} = \sum_{k=0}^\infty \binom{-n}k x^k, \\
\binom{-n}k = \frac{(-n)(-n-1)\ldots(-n-k+1)}{k!} 
= (-1)^k \binom{n+k-1}k
\label{(-n k)}
\end{gather}

\begin{multline}
\int_0^{2\pi}
\frac {e^{iar e^{i\phi} } e^{- i(n-1)\phi} \ud\phi}
      {(1 + \frac{r}{2i} e^{i\phi})^n} 
\\
=
\int_0^{2\pi}
\sum_{l=0}^\infty \frac {(iare^{i\phi})^l}{l!}
\sum_{k=0}^\infty 
    % {-n \chose k} % amsmath doesn't allow \chose -- Lvovsky.
    \binom{-n}k
    \left(\frac{r}{2i} e^{i\phi}\right)^k
e^{-i(n-1)\phi}
\ud \phi = \\
=
2\pi 
\sum_{k=0}^{n-1}
\binom{-n}k \left(\frac r{2i}\right)^k \frac{(iar)^{n-1-k}}{(n-1-k)!}
\\
=
2\pi r^{n-1}
\sum_{k=0}^{n-1} (-1)^k \binom{n+k-1}{k}
\frac{i^{n-1-2k}a^{n-1-k}}{2^k(n-1-k)!}
\\
=
\frac{2\pi(ir)^{n-1}}{2^{n-1}} 
\sum_{k=0}^{n-1} \binom{n+k-1}{n-1} \frac{(2a)^{n-1-k}}{(n-1-k)!}
\\
= \frac{2\pi(ir)^{n-1}}{2^{n-1}} 
\sum_{k=0}^{n-1} \binom{2n-2-k}{n-1} \frac{(2a)^{k}}{k!}
\end{multline}

Finally, we obtain

\begin{equation}
\int_0^\infty 
    \frac{\cos(a x)}{(x^2 + 1)^{n}}
\ud x
=
\frac{\pi e^{-a}}{2^{2n-1}}
\sum_{k=0}^{n-1} \binom{2n-2-k}{n-1} \frac{(2a)^{k}}{k!}
\label{E_cos_int_exact}
\end{equation}

\paragraph{\fbox{$n=0$}}

As in the previous case% $n\neq0$
, we transform

\begin{equation}
\int_0^{\infty}
\frac {x\sin(ax)}{x^2+1}
\ud x = 
\frac 12 \Im \int_{-\infty}^{\infty}
\frac {xe^{iax}}{x^2+1}
\ud x
\end{equation}

As earlier, we close the contour of integration in the upper half-plane, 
deform the contour to the pole $+i$ and parameterise the integration variable
$z = i + re^{i\phi}$.

\begin{align}
\int_0^{2\pi}
\frac{(i+re^{i\phi})e^{-a + iar e^{i\phi}} ir e^{i\phi}}
     {2ire^{i\phi} + r^2 e^{2i\phi}}
\ud \phi 
& =
e^{-a}
\int_0^{2\pi}
    \frac{(i+re^{i\phi}) e^{iar e^{i\phi}} }
         {2 + \frac ri e^{i\phi}}
\ud \phi
\label{GR_int_n_eq_0}
\end{align}

The integrand is analytic on $[0,2\pi]$ and thus 
can be expressed as a series of non-negative powers of 
$e^{i\phi}$, from which 
only the zeroth term 
contributes to the integral and gives $\pi i$. Therefore

\begin{equation*}
\int_0^{\infty}
\frac {x\sin(ax)}{x^2+1}
\ud x = 
\frac \pi2 e^{-a}.
\end{equation*}

%% file: exp_cdf.tex
\subsection
{
CDF\texorpdfstring{$_E(\cos\theta(\mathbf r_n))$}{(cosθ(rₙ))}
}

The original problem in~\cite{Chooz99}
was to find the half of the opening angle of the cone
whose origin is at
(0,0,0) and the axis is the true direction, 
within which the given confidence level $cl$ (e.g. 68\%)
of events are contained. 

When we work in spherical coordinates $(r,\phi,\theta)$, 
where $\theta$ is the angle to the true direction $\mathbf{r}_0$,
we can compute the $\mathrm{CDF}$ as a function of $\theta$.
Then the confidence interval $\theta_{cl}$ is $\mathrm{CDF}^{-1}(cl)$. 
This can be calculated numerically if we know the CDF.

In this subsection we calculate the cumulative distribution function 
of the polar angle $\theta$ of the statistic $\mathbf r_n$.
For more compact formulae, we calculate the CDF as a function of
$\cos \theta$.

Changing to spherical coordinates in~\ref{E_n_full} and integrating on 
$\phi$ and $r$, 

\begin{multline}
\mathrm{CDF}_E(\cos\theta) 
=
\frac{n^3}{l^3}
\int_0^\infty r_n^2\,\ud r_n 
\int_{\cos\theta}^1
    \frac{e^{-\frac nl \sqrt{r_n^2 + r_0^2 - 2 r_n r_0 \cos\theta'}}}
         {2^{4n-2}(2n-1)!}
\cdot
\\
\cdot \sum_{k=0}^{2n-2} 
\frac{(4n-4-k)!(2 \frac nl \sqrt{r_n^2 + r_0^2 - 2 r_n r_0 \cos\theta'})^k}
     {k!(2n-2-k)!}
\,\ud \cos\theta'
\label{CDF_E_begin}
% eqnarray doesn't work with showkeys!
\end{multline}

\subsubsection*{Integral on $\cos\theta'$}
For shorter 
notation we define

\begin{equation}
\left(
\begin{array}{rcl}
a & = & r_n^2 + r_0^2 \\
b & = & 2 r_n r_0 \\
c & = & \frac nl
\end{array}
\right)
\label{abc}
\end{equation}

We assume $b\neq0$, that is we exclude the point $r=0$ from the integration.
We calculate

\begin{align}
\int_x^1 
e^{-c\sqrt{a - bx'}} (c\sqrt{a - bx'})^k
\ud x' 
& =
\nonumber
\\
\left(
z = c\sqrt{a - bx'}
\atop
\ud x' = -\frac{2z\ud z}{bc^2}
\right)
& =
- \frac{2}{bc^2}
\int_{c\sqrt{a - bx}}^{c\sqrt{a - b}}
e^{-z}z^{k+1}\,\ud z
\label{int_sqrt}
\end{align}

\begin{equation}
\int x^k e^{-x}\,\ud x = -x^k e^{-x} + k \int x^{k-1} e^{-x} \,\ud x = 
-e^{-x}k! \sum_{i=0}^k \frac{x^i}{i!}
\label{int_e-x_xk}
\end{equation}

Combining \ref{abc}, \ref{int_sqrt}, \ref{int_e-x_xk} 

\begin{gather*}
\int_{\cos\theta}^1
e^{-\frac nl \sqrt{r_n^2 + r_0^2 - 2 r_n r_0 \cos\theta'}}
\left(\frac nl \sqrt{r_n^2 + r_0^2 - 2 r_n r_0 \cos\theta'}\right)^k
\,\ud \cos\theta'
=
\nonumber
\\
=
\frac{l^2}{n^2 r_n r_0} e^{-z}(k+1)!
\sum_{i=0}^{k+1}
\left.
\frac{z^{i}}{i!}
\right|_{\frac nl \sqrt{r_n^2 + r_0^2 - 2 r_n r_0 \cos \theta}}
       ^{\frac nl \sqrt{r_n^2 + r_0^2 - 2 r_n r_0}}
=
\nonumber 
\\
=
\frac{l^2(k+1)!}{n^2 r_0} 
\sum_{i=0}^{k+1} 
\frac{1}{i!}
\Big(
\nonumber
\hspace{7cm}
\nonumber 
\\
\qquad \quad
\frac 1{r_n} 
e^{-\frac nl |r_n - r_0|} 
\left(\frac nl |r_n - r_0|\right)^{i} -
\label{int_cos_upper}
\addtocounter{equation}{1}
\tag{\textrm{\theequation}}
\\
\shoveright
- 
\frac 1{r_n} 
e^{- \frac nl \sqrt{r_n^2 + r_0^2 - 2 r_n r_0 \cos \theta}}
\left(\frac nl \sqrt{r_n^2 + r_0^2 - 2 r_n r_0 \cos \theta}\right)^{i}
\addtocounter{equation}{1}
\tag{\textrm{\theequation}}
\label{int_cos_lower}
\\
\qquad 
\hspace{-3cm}
\Big)
.
\addtocounter{equation}{1}
\tag{\textrm{\theequation}}
\label{CDF_E_int_cos}
\end{gather*}
% split doesn't allow \tag .

\subsubsection*
{Integral of \ref{int_cos_upper} over $r_n$}
\label{sss_int_cos_upper}

Integrating \ref{int_cos_upper}, we multiply it by $r_n^2$ from the Jacobean,
and expand the modulus

\begin{align}
\int_0^\infty 
r_n e^{-\frac nl |r_n - r_0|}
\left(\frac nl |r_n - r_0| \right)^i
& \ud r_n 
=
\label{int_cos_upper_dr}
\\
&
\int_0^{r_0}
r_n e^{-\frac nl (r_0 - r_n)}
\left(\frac nl (r_0 - r_n)\right)^i
\ud r_n 
\label{int_module_zero}
\\
& 
+ 
\int_{r_0}^\infty
r_n e^{-\frac nl (r_n - r_0)}
\left(\frac nl (r_n - r_0)\right)^i
\ud r_n 
\label{int_module_infty}
% aligned doesn't support multiple labels.
\end{align}

The integral from $0$ to $r_0$
is easily calculated using \ref{int_e-x_xk}:

\begin{multline}
\textrm{\ref{int_module_zero}} 
=
\left(
\begin{array}{c}
\frac nl \left(r_0 - r_n\right) = x, \\
r_n = r_0 - \frac ln x , \\
\ud r_n = - \frac ln \ud x 
\end{array}
\right)
=
- \int_{\frac nl r_0}^0
\frac ln \left(r_0 - \frac ln x\right) e^{-x}x^i
\,\ud x
\stackrel{\left( \textrm{ \ref{int_e-x_xk} } \right)}{=} 
\\
=
\left.
\frac ln r_0 e^{-x}
i! \sum_{j=0}^{i} \frac{x^j}{j!}
\right|_{\frac nl r_0}^0
- 
\frac{l^2}{n^2} 
\int_0^{\frac nl r_0} x^{i+1}e^{-x} \ud x
= 
\\
=
\frac ln r_0 i! - \frac ln r_0 e^{-\frac nl r_0}
i! \sum_{j=0}^{i} \frac{\left(\frac nl r_0\right)^j}{j!}
-
\frac{l^2}{n^2} 
\int_0^{\frac nl r_0} x^{i+1}e^{-x} \ud x
\label{int_module_zero_sol}
\end{multline}

We have retained the last integral, for it 
%will be 
is
useful in what follows.

The integral from $r_0$ to infinity is taken similarly:

\begin{multline}
\mbox{\ref{int_module_infty}}
=
\frac ln \int_{r_0}^\infty 
  e^{-\frac nl(r_n - r_0)}\left(\frac nl(r_n - r_0)\right)^{i+1}
\ud r_n
\\
+ r_0
\int_{r_0}^\infty 
  e^{-\frac nl(r_n - r_0)}\left(\frac nl(r_n - r_0)\right)^i
\ud r_n
=
\left( \frac ln \right)^2 (i+1)! + \frac ln r_0 i!
\label{int_module_infty_sol}
\end{multline}

Adding \ref{int_module_zero_sol} and \ref{int_module_infty_sol}, we obtain

\begin{align}
\mbox{\ref{int_cos_upper_dr}} & =
2\,i! \frac ln r_0 
- \frac ln r_0 e^{-\frac nlr_0} i!
\sum_{j=0}^i \frac{\left(\frac nl r_0\right)^j}{j!}
+ 
(i+1)!\frac{l^2}{n^2} 
-
\frac{l^2}{n^2} 
\int_0^{\frac nl r_0} x^{i+1}e^{-x} \ud x
\label{int_cos_upper_sol}
\end{align}

In case of a large number of events $n$ or, more precisely, 
when $\frac nl r_0 \gg 1$,
the exponent $e^{-\frac nl r_0}$ is much smaller than any power of $\frac nl r_0$, 
and 

\begin{equation}
\mbox{\ref{int_cos_upper_dr}} 
\simeq 
2\, i!\,\frac ln \,r_0
.
\end{equation}

This corresponds to the case when most of the integral \ref{int_cos_upper_dr}
is accumulated in a neighbourhood of $r_n = r_0$.

\subsubsection*{Integral of \ref{int_cos_lower} over $r_n$}
\label{sss_int_cos_lower}
In this subsubsection we calculate the integral

\begin{equation}
\int_0^\infty
  r_n e^{-\frac nl \sqrt{r_n^2 + r_0^2 - 2 r_n r_0 \cos \theta} } 
  \left( \frac nl \sqrt{r_n^2 + r_0^2 - 2 r_n r_0 \cos \theta} \right)^i 
\,\ud r_n 
\label{int_cos_sqrt_int}
.
\end{equation}

We introduce

\begin{equation*}
x = \frac nl \sqrt{r_n^2 + r_0^2 - 2 r_n r_0 \cos \theta}, 
\end{equation*}

then we can rewrite

\begin{equation*}
x^2 \left(\frac ln\right)^2
=
(r_n - r_0 \cos\theta)^2 + r_0^2(1-\cos^2\theta) 
\end{equation*}

Change from $r_n$ to $x$ is a change of coordinates 
if $r_n$ is uniquely defined through $x$ and vice versa. 
Therefore we should separately consider the regions
$r_n \geq r_0 \cos\theta$ and $r_n < r_0 \cos\theta$. 
The point $r_n = r_0 \cos\theta$ on a line of integration corresponds 
to the maximum of the pdf on that line 
(this is the nearest point on the line to the the mode of the distribution).

\paragraph{
\fbox{\fbox{
    $\cos\theta \geq 0$
}}
}

\paragraph{
\fbox{
    $r_n \geq r_0 \cos\theta$
}
}

\begin{eqnarray*}
& & r_n = r_0 \cos\theta + \sqrt{\frac{l^2}{n^2}x^2 - r_0^2(1 - \cos^2\theta)}
\nonumber
\\
& & 
\ud r_n = 
\frac{\frac{l^2}{n^2} x \ud x}
     {\sqrt{\frac{l^2}{n^2}x^2 - r_0^2(1 - \cos^2\theta)}}
=
\frac{\frac{l}{n} x \ud x}
     {\sqrt{x^2 - \frac{n^2}{l^2}r_0^2(1 - \cos^2\theta)}}
\end{eqnarray*}

\begin{eqnarray}
\int_{r_0\cos\theta}^\infty
r_n 
e^{-\frac nl 
%\sqrt{(r_n - r_0 \cos\theta)^2 + r_0^2(1-\cos^2\theta)}
% too long!
\sqrt{r_n^2 + r_0^2 - 2 r_n r_0 \cos \theta}
} 
\left(\frac nl 
\sqrt{r_n^2 + r_0^2 - 2 r_n r_0 \cos \theta}
\right)^i 
\,\ud r_n =
\nonumber \\
=
\frac ln 
r_0 
\cos\theta
\int_{\frac nl r_0 \sqrt{1 - \cos^2\theta}}^\infty
\frac{x^{i+1}}
     {\sqrt{x^2 - \frac{n^2}{l^2}r_0^2(1 - \cos^2\theta)}}
e^{-x}
\,\ud x
\ +
\label{r0>costheta_1}
& &
\\
\lefteqn{
\phantom{=}
+ 
\frac {l^2}{n^2} 
\int_{\frac nl r_0 \sqrt{1 - \cos^2\theta}}^\infty
x^{i+1} e^{-x}
\,\ud x
}
\phantom{
=
\int_{\frac  nl r_0 \sqrt{1 - \cos^2\theta}}^\infty
r_0 \cos\theta
\frac{\frac ln x^{i+1}}
     {\sqrt{x^2 - \frac{n^2}{l^2}r_0^2(1 - \cos^2\theta)}}
e^{-x}
\,\ud x
+
}
& &
\label{gamma_cos}
\end{eqnarray}

The integral \ref{r0>costheta_1} converges 
at the lower limit
for $\cos\theta \neq 1$ 
because
the 
integral 
$\int_a^b \frac{\ud x}{\sqrt{x^2 - a^2}} = 
\int_a^b \frac{\ud x}{\sqrt{(x-a)(x+a)}}$
converges.
The integral

\begin{equation*}
\int_a^\infty \frac{x^{i+1}e^{-x}\ud x}{\sqrt{x^2 - a^2}} 
\stackrel{(x=a\mathrm{ch}y)}{=} 
a^{i+1}\int_0^\infty \mathrm{ch}^{i+1}\!y\ e^{-a\mathrm{ch}y}\ud y
\end{equation*}

can be expressed through 
the modified Bessel function $K_\nu$ (8.407 in~\cite{GradRyzh})
using the formula 3.547(4) from~\cite{GradRyzh}:

\begin{equation*}
\int_0^\infty \mathrm{exp}\left(-\beta\mathrm{cosh}x\right)
\mathrm{cosh}\left(\gamma x\right) \ud x = K_\gamma\left(\beta\right)
\qquad \qquad \quad
\left[\Re\beta > 0\right]
\end{equation*}

since ch$^nx$ can be expressed as a sum of ch($kx$) using 1.320(6) and 1.320(8) 
from~\cite{GradRyzh}.

\paragraph{
\fbox{
$ 0 \leq r_n \leq r_0 \cos\theta$
}
}

\begin{eqnarray*}
\sqrt{\frac{l^2}{n^2}x^2 - r_0^2(1 - \cos^2\theta)} = r_0 \cos\theta - r_n, \\
r_n = r_0\cos\theta - \sqrt{\frac{l^2}{n^2}x^2 - r_0^2(1 - \cos^2\theta)}
\end{eqnarray*}

The limits $r_n|_0^{r_0\cos\theta}$ 
are converted to
$x|_{\frac  nl r_0}^{\frac nl r_0 \sqrt{1-\cos^2\theta}}$. 
The differential $\ud r_n$ is the same as in the previous case 
$r_n \geq r_0\cos\theta$ except for the negative sign,
which we omit changing the upper and the lower limits of the integration.

\begin{multline}
\int_0^{r_0\cos\theta}
    r_n 
    e^{-\frac nl \sqrt{r_n^2 + r_0^2 - 2 r_n r_0 \cos \theta}
      }
    \left(
        \frac nl \sqrt{r_n^2 + r_0^2 - 2 r_n r_0 \cos \theta}
    \right)^i 
\,\ud r_n 
=
\\
=
\int_{\frac nl r_0 \sqrt{1-\cos^2\theta}}^{\frac  nl r_0}
    \left(
        r_0\cos\theta - \sqrt{\frac{l^2}{n^2}x^2 - r_0^2(1-\cos^2\theta)}
    \right)
    \frac{
         e^{-x} \frac ln x^{i+1} \,\ud x
    }
    {\sqrt{x^2 - \frac{n^2}{l^2}r_0^2(1 - \cos^2\theta)}}
=
\\
\begin{split}
=
&
\; 
\frac ln r_0\cos\theta
\int_{\frac nlr_0 \sqrt{1-\cos^2\theta}}^{\frac nlr_0}
    \frac{x^{i+1}}
         {\sqrt{x^2 - \frac{n^2}{l^2}r_0^2(1 - \cos^2\theta)}}
    e^{-x}
\,\ud x 
\\
& -
\frac {l^2}{n^2} 
\int_{\frac nlr_0 \sqrt{1-\cos^2\theta}}^{\frac nl r_0}
    x^{i+1} e^{-x}
\,\ud x
.
\end{split}
\label{r0_lt_costheta}
\end{multline}

Adding \ref{r0>costheta_1} and \ref{gamma_cos} to 
\ref{r0_lt_costheta} 
gives

\begin{gather}
\textrm{\ref{int_cos_sqrt_int}}
\stackrel{1 > \cos\theta \geq 0}{=}
\label{E_CDF_int_cos_gt_0}
\\
=
\frac ln r_0\cos\theta
\int_{\frac  nl r_0}^\infty
    \frac{x^{i+1}}
         {\sqrt{x^2 - \frac{n^2}{l^2}r_0^2(1 - \cos^2\theta)}}
    e^{-x}
\,\ud x 
+
\frac {l^2}{n^2}
\int_{\frac  nl r_0}^\infty
    x^{i+1} e^{-x}
\,\ud x
+
\\
+
2 \frac ln r_0\cos\theta
\int_{\frac nl r_0 \sqrt{1-\cos^2\theta}}^{\frac  nl r_0}
    \frac{x^{i+1}}
         {\sqrt{x^2 - \frac{n^2}{l^2}r_0^2(1 - \cos^2\theta)}}
    e^{-x}
\,\ud x 
\label{int_cos_gt_0_add}
.
\end{gather}

\paragraph{
\fbox{\fbox{
    $\cos\theta < 0$
}}
}

In this case $r_n$ is always bigger than $r_0 \cos\theta$
and the integral becomes as in the case of \ref{r0>costheta_1}, 
\ref{gamma_cos}.

The lower limit is changed from $r_n = 0$ to $x = \frac nl r_0$,

\begin{multline}
\textrm{\ref{int_cos_sqrt_int}}
\stackrel{\cos\theta < 0}{=}
\frac ln 
r_0 
\cos\theta
\int_{\frac nl r_0}^\infty
\frac{x^{i+1}}
     {\sqrt{x^2 - \frac{n^2}{l^2}r_0^2(1 - \cos^2\theta)}}
e^{-x}
\,\ud x
+ 
\frac {l^2}{n^2}
\int_{\frac nl r_0}^\infty
    x^{i+1} e^{-x}
\,\ud x.
\label{int_cos_lt_0}
\end{multline}

Note that the only difference between \ref{int_cos_lt_0} and 
\ref{E_CDF_int_cos_gt_0} is
\ref{int_cos_gt_0_add}.

We can combine the results for 
$\cos\theta < 0$ 
and for 
$\cos\theta \geq 0$
using the Heaviside step function:

\begin{equation}
\Theta(x) = 
\begin{cases}
1 & x \geq 0, \\
0 & x < 0.
\end{cases}
\label{Theta_Heaviside}
\end{equation}

\begin{multline}
\textrm{\ref{int_cos_sqrt_int}}
=
\frac ln 
r_0 
\cos\theta
\int_{\frac nl r_0}^\infty
\frac{x^{i+1}}
     {\sqrt{x^2 - \frac{n^2}{l^2}r_0^2(1 - \cos^2\theta)}}
e^{-x}
\,\ud x
+
\frac {l^2}{n^2}
\int_{\frac nl r_0}^\infty
    x^{i+1} e^{-x}
\,\ud x
\\
+ 
2 
\Theta(\cos\theta)
\frac ln r_0\cos\theta
\int_{\frac nl r_0 \sqrt{1-\cos^2\theta}}^{\frac  nl r_0}
    \frac{x^{i+1}}
         {\sqrt{x^2 - \frac{n^2}{l^2}r_0^2(1 - \cos^2\theta)}}
    e^{-x}
\,\ud x
\label{int_cos_theta_sqrt_full}
\end{multline}

\subsubsection*{CDF($\cos\theta(\mathbf r_n))$}
Combining the calculations for the CDF$(\cos\theta)$, 

\begin{gather}
\mathrm{CDF}(\cos\theta) 
\stackrel{\ref{CDF_E_begin},\ref{CDF_E_int_cos}}{=}
\frac nl \frac 1{r_0}
\sum_{k=0}^{2n-2}
\frac{(4n-4-k)!2^k(k+1)!}
 {k!(2n-2-k)!}
\int_0^{\infty}r_n \ud r_n
\sum_{i=0}^{k+1} \frac 1{i!}
\Bigg(
\nonumber
\\ 
e^{-\frac nl |r_n - r_0|}\left(\frac nl |r_n - r_0|\right)^i
- e^{-\frac nl\sqrt{r_n^2 + r_0^2 - 2r_nr_0\cos\theta}}
\left(\frac nl \sqrt{r_n^2 + r_0^2 - 2 r_n r_0 \cos\theta}\right)^i
\nonumber
\\
\Bigg)
\stackrel{\ref{int_cos_upper_sol},\ref{int_cos_theta_sqrt_full}}{=}
\frac nl \frac 1{r_0}
\sum_{k=0}^{2n-2}
\frac{(4n-4-k)!2^k(k+1)}
 {(2n-2-k)!}
\sum_{i=0}^{k+1} \frac 1{i!}
\Bigg(
\nonumber
\\
% |r_n - r_0| part.
2\,i! \frac ln r_0 
- \frac ln r_0 e^{-\frac nlr_0} i!
\sum_{j=0}^i \frac{\left(\frac nl r_0\right)^j}{j!}
+
(i+1)!\frac{l^2}{n^2} 
-
\frac{l^2}{n^2} 
\int_0^{\frac nl r_0} x^{i+1}e^{-x} \ud x
\label{int_cos_upper_sol_here}
\\
% \sqrt{..\cos\theta..} part.
% copied from above
-
\frac ln 
r_0 
\cos\theta
\int_{\frac nl r_0}^\infty
\frac{x^{i+1}}
     {\sqrt{x^2 - \frac{n^2}{l^2}r_0^2(1 - \cos^2\theta)}}
e^{-x}
\,\ud x
-
\frac {l^2}{n^2}
\int_{\frac nl r_0}^\infty
    x^{i+1} e^{-x}
\,\ud x
% end copied
\label{int_cos_sqrt_here}
\\
- 
2 
\Theta(\cos\theta)
\frac ln r_0\cos\theta
\int_{\frac nl r_0 \sqrt{1-\cos^2\theta}}^{\frac  nl r_0}
    \frac{x^{i+1}}
         {\sqrt{x^2 - \frac{n^2}{l^2}r_0^2(1 - \cos^2\theta)}}
    e^{-x}
\,\ud x
\nonumber
\Bigg)
\end{gather}

The line \ref{int_cos_upper_sol_here} corresponds to \ref{int_cos_upper},
while \ref{int_cos_sqrt_here} and below come from \ref{int_cos_lower}.

The last terms in \ref{int_cos_upper_sol_here} and \ref{int_cos_sqrt_here} 
sum up to $-\frac{l^2}{n^2}\Gamma(i+2)$ and cancel with $(i+1)!\frac{l^2}{n^2}$
in \ref{int_cos_upper_sol_here}.  
Thus we obtain the final answer:

\begin{multline}
\mathrm{CDF}_E(\cos\theta(\mathbf{r}_n, \mathbf{r}_0)) =
\sum_{k=0}^{2n-2}
\frac{(4n-4-k)!2^k(k+1)}
 {(2n-2-k)!}
\Bigg(
2(k+2)
\\
- e^{-\frac nlr_0}
\sum_{i=0}^{k+1} 
(k+2-i)\frac{\left(\frac nl r_0\right)^i}{i!}
-
\cos\theta
\int_{\frac nl r_0}^\infty
\frac{\sum_{i=0}^{k+1} \frac{x^{i+1}}{i!}}
     {\sqrt{x^2 - \frac{n^2}{l^2}r_0^2(1 - \cos^2\theta)}}
e^{-x}
\,\ud x
\\
- 
2 
\Theta(\cos\theta)
\cos\theta
\int_{\frac nl r_0 \sqrt{1-\cos^2\theta}}^{\frac  nl r_0}
\frac{\sum_{i=0}^{k+1} \frac{x^{i+1}}{i!}}
         {\sqrt{x^2 - \frac{n^2}{l^2}r_0^2(1 - \cos^2\theta)}}
    e^{-x}
\,\ud x
\Bigg)
\label{E_CDF_cos_theta_final}
\end{multline}

The first term in \ref{E_CDF_cos_theta_final} is equal to $2^{4n-2}(2n-1)!$
due to \ref{normfact}.

%% file: exp_cdf_approx.tex
\subsection
{
Approximation of 
\texorpdfstring{$E_n$}{Eₙ} 
and 
\texorpdfstring{$\theta(cl)$}{θ(cl)}
for $n$ large
}

As in~\ref{exp_integral_GR}, we use 

\begin{equation}
a_E = \frac nl |r_n - r_0|
\label{a_E}
\end{equation}

for briefer notation. 
We also introduce
\footnote{the subscript E and the dependence F(a) are omitted throughout this section}

\begin{equation}
F_n = \frac{n^3}a \int_0^{\infty} \frac{x\sin(ax)}{(1+x^2)^n} \ud x,
\label{E_F_n}
\end{equation}

then the pdf \ref{E_n} can be expressed as

\begin{equation}
E_n = \frac 1{16\pi^2l^3} F_{2n}.
\label{E_n_F_n}
\end{equation}

Let also 

\begin{equation}
I_n = \int_0^{\infty} \frac{\cos(ax)}{(1+x^2)^n} \ud x,
\label{E_I_n}
\end{equation}

then due to \ref{GR_cos_int}

\begin{equation}
F_n = \frac{n^3}{2(n-1)} I_{n-1}.
\label{E_F_n_I_n}
\end{equation}

Even though we have the exact formula \ref{GR_int} for $I_n$, 
we need to find a simple estimate for that when $n$ is large.

The function $(1+x^2)^{-n}$ has its maximum at $x=0$ and rapidly decreases to zero
when $x$ and $n$ increase.
Thus the integral 
\ref{E_I_n}
can accumulate most of its value near $x=0$.
For that to happen, $\cos(ax)$ should be positive when $(1+x^2)^n$ is large enough.

Small $a$ (large $I_n(a)$) correspond to the maximum of the pdf, 
while large $a$ (small $I_n(a)$) correspond to the tail of the distribution.
If we are interested in confidence levels neither too close to $1$, 
nor too close to $0$, 
we should explore the region where $I_n(a)$ takes intermediate values.

%To estimate the confidence interval for the given confidence level
%we need to understand for what parameters of $a$ the function \ref{GR_int}
%is neither too large, nor too small, 
%which would correspond to too small or too large confidence intervals.
%The confidence level can be any on $(0,1)$, but it should not depend on $n$,
%and the accuracy of the approximation for the $cl$ very close to $0$ or $1$
%will be lower than that of other cl-s, but will increase when increasing $n$.

\begin{figure}
\subfigure[Smaller $a$]{
    \label{fig:cos_over_pow_small_param}
    \includegraphics%[width=0.3\textwidth]
        {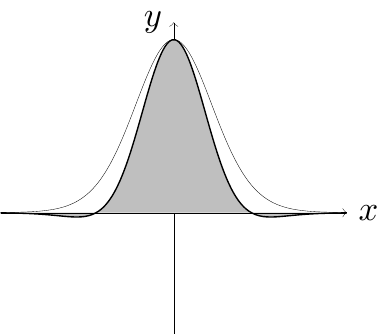}
}
\subfigure[Larger $a$]{
    \label{fig:cos_over_pow_large_param}
    \includegraphics%[width=0.3\textwidth]
        {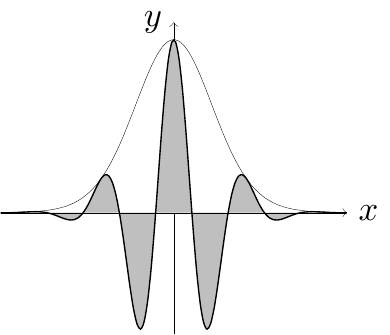}
}
\subfigure[$a_*$]{
    \label{fig:cos_over_pow_special_param}
    \includegraphics%[width=0.3\textwidth]
        {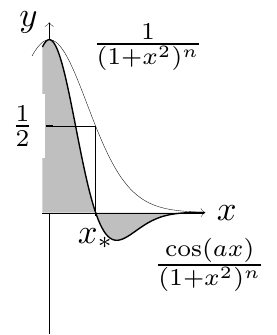}
}
\caption{$\frac{\cos(ax)}{(1+x^2)^n}$ for different $a$ values. The upper curve represents $\frac 1{(1+x^2)^n}$.}
\label{fig:cos_over_pow}
\end{figure}

On figure \ref{fig:cos_over_pow} three different cases for the parameter $a$ are shown.
On \ref{fig:cos_over_pow_small_param} $a$ is small, 
and the integral \ref{E_I_n} is close to its maximum 
(as if $a=0, \cos(ax) = 1$).
On \ref{fig:cos_over_pow_large_param} $a$ is large, the $\cos(ax)$ oscillates fast 
%compared to the decrease of 
and the integral \ref{E_I_n} is small. 
To estimate the parameter of interest $a_*$, 
we define it such that the first zero of
$\cos(ax)$ is when $(1+x^2)^{-n}$ is neither too large, nor too small.
For an estimate
we set it to $\frac 12$, fig.~\ref{fig:cos_over_pow_special_param}.

Solving $\frac 1{(1 + x^2)^n} = \frac 12$ gives
$x_*^2 = 2^{\frac 1n} - 1$ and

\begin{equation}
x_* \approx \frac{\sqrt{\mathrm{ln}(2)}}{\sqrt n}.
\end{equation}

$a_*$ such that $\cos(a_*x_*) = 0$ is equal to 
$a_* = \frac\pi{2x_*} \approx \frac{\pi\sqrt n}{2\sqrt{\mathrm{ln}(2)}}$.

We can change $2$ from the example to another number 
$A$
, then $a_*$ changes to 

\begin{equation}
a_* \approx \frac{\pi\sqrt n}{2\sqrt{\mathrm{ln}(A)}}
,
\label{a_star}
\end{equation}

which depends very weakly on $A$ as $A$ grows. Therefore 
$a$ of interest for our problem is less than or of the order of $\sqrt n$.

The derivative w.r.t. $a$ of the integrand of $I_n$, 
$\left| \frac{-x \sin(ax)}{(1+x^2)^n} \right| < \frac x{(1+x^2)^n}$.
The integral $\int_0^{\infty} \frac x{(1+x^2)^n}\ud x$ converges for $n>1$,
therefore, according to the well-known theorem \cite{wIntegral_parametrique},
we can differentiate \ref{E_F_n_I_n} on the parameter:

\begin{equation}
\left(I_n\right)'_a = 
\int_0^{\infty} \frac{-x\sin(ax)}{(1+x^2)^n} \ud x 
\stackrel{\textrm{\ref{GR_cos_int}}}{=}
-\frac{a}{2(n-1)}I_{n-1}.
\label{E_I_n_stroke}
\end{equation}

Now we derive a recursion formula for $I_n$.

\begin{gather*}
I_n = \int_0^\infty \frac{(1+x^2)\cos(ax)}{(1+x^2)^{n+1}} \ud x
= 
I_{n+1} + \int_0^\infty \frac{x^2\cos(ax)}{(1+x^2)^{n+1}} \ud x,
\\
\int_0^\infty \frac{x^2\cos(ax)}{(1+x^2)^{n+1}} \ud x
=
-\left. \frac 1{2n} \frac 1{(1+x^2)^n} x\cos(ax) \right|_0^\infty + 
\\
+ \frac 1{2n} \int_0^\infty \frac{\cos(ax) - ax\sin(ax)}{(1+x^2)^n} \ud x
=
\frac 1{2n}I_n - \frac{a^2}{4n(n-1)}I_{n-1},
\end{gather*}

therefore

\begin{equation}
I_{n+1} = \left(1 - \frac 1{2n}\right) I_n + \frac{a^2}{4n(n-1)} I_{n-1}.
\label{E_I_n_rec}
\end{equation}

\begin{equation}
I_{n+1} = I_n + 
O\left(n^{-1}\right)I_n
\label{E_I_n_rec_approx}
,\textrm{ for $a \lesssim \sqrt n$ }
,
\end{equation}

and substituting \ref{E_I_n_rec_approx} into \ref{E_I_n_stroke}, 
we can solve the differential equation up to the terms of the order of $\frac 1n$:

\begin{gather}
\frac {\ud I_n(a)}{\ud a} 
= -\frac a{2n} I_n \left(1 + \mathit O\left(n^{-1}\right)\right),
\nonumber
\\
I_n(a) = I_n(0) e^{-\frac {a^2}{4n} \left(1+\mathit O\left(n^{-1}\right)\right)}
=
I_n(0) e^{-\frac {a^2}{4n}} \left(1 + \mathit O\left(n^{-1}\right)\right)
.
\label{E_I_n_approx}
\end{gather}

$I_n(0)$ can be found from the exact formula \ref{E_cos_int_exact},

\begin{equation}
I_n(0) = \frac{\pi}{2^{2n-1}} \frac{(2n-2)!}{((n-1)!)^2}
.
\label{E_I_n_0}
\end{equation}

We can approximate $I_n(0)$ for large $n$ using Stirling's formula 
\cite{StirlingsFormula}

\begin{equation}
n! = \sqrt{2\pi n} \left(\frac ne\right)^n 
\left(1 + \mathit O\left(\frac 1n\right)
\right),
\label{Stirling}
\end{equation}

then

\begin{equation*}
\frac{(2n-2)!}{((n-1)!)^2} 
\approx
\frac{\sqrt 2}{\sqrt{2\pi(n-1)}} \frac{(2n-2)^{2n-2}}{(n-1)^{2n-2}}
=
\frac{2^{2n-2}}{\sqrt \pi \sqrt{n-1}},
\end{equation*}

therefore

\begin{equation}
I_n(0) = \frac{\sqrt \pi}{2\sqrt n} + \mathit O(n^{-\frac 32}).
\label{E_I_n_0_approx}
\end{equation},

and

\begin{gather}
I_n 
\stackrel{\ref{E_I_n_approx},\ref{E_I_n_0_approx}}{=}
\frac{\sqrt \pi}{2\sqrt n} e^{-\frac{a^2}{4n}}
        \left(1 + \mathit O(n^{-1})\right)
\label{I_n_approx_fin}
, 
\\
F_n 
\stackrel{\ref{E_F_n_I_n},\ref{I_n_approx_fin}}{=}
\frac{n^{\frac 32}\sqrt \pi}4 e^{-\frac{a^2}{4n}}
         \left(1 + \mathit O(n^{-1})\right)
, 
\label{F_n_approx_fin}
\\
E_n 
\stackrel{\ref{E_n_F_n},\ref{F_n_approx_fin}}{=}
\frac{(2n)^{\frac 32}}{64\pi^{\frac 32}l^3}
e^{-\frac{a^2}{8n}}
         \left(1 + \mathit O(n^{-1})\right)
\label{E_n_approx_fin}
.
\end{gather}

Substituting $a_E$ from \ref{a_E}, we express $E_n$ through the original parameters:
\begin{equation}
E_n(\mathbf r_n) 
=
\frac{n^{\frac 32}}{(2\pi)^\frac{3}{2}8l^3}
e^{-\frac{n\left|r_n - r_0\right|^2}{8l^2}}(1 + \mathit O(n^{-1}))
,
\label{E_n_r_n_approx_fin}
\end{equation}

and obtain in the leading order the normal distribution \ref{gauss_d} with

\begin{equation}
\sigma_{E,n} = \frac{2l}{\sqrt n}
\label{E_sigma}
,
\end{equation}

which corresponds to the convolution of $n$ normal distributions with
$\sigma_{E,1} = 2l$. In the limit of the number of events very large 
($\sqrt{n}\frac{r_0}{2l} \gg 1 \textrm{ and } \theta \ll 1$), 
one can apply the formula \ref{G_theta_large_n_simple}
to estimate a confidence interval $\theta$:

\begin{equation}
\theta_E \approx \frac{2l\sqrt{-2\mathrm{ln}(1-cl)}}{r_0\sqrt n}
\label{E_theta_large_n}
.
\end{equation}

%% file: gauss.tex
\section{Normal distribution}
\label{normal_distribution}

\subsection
{Introduction. Distribution of the sample mean 
\texorpdfstring{$G_n$}{Gₙ}}
A one-dimensional normal (or Gaussian) distribution has the pdf

\begin{equation}
g(x) = \frac 1{\sqrt{2\pi\sigma^2}} e^{-\frac{(x-x_0)^2}{2\sigma^2}}.
\end{equation}

With this definition the variance $\mathrm E[(x-x_0)^2] = \sigma^2$.
For $d$ dimensions the spherically symmetric multivariate normal distribution is

\begin{equation}
g(\mathbf r) = 
\frac 1{(2\pi\sigma^2)^{\frac d2}} 
               e^{-\frac{(\mathbf r- \mathbf r_0)^2}{2\sigma^2}}
\label{gauss_d}
\end{equation}

\paragraph{The Fourier transform} 
of the multivariate normal distribution (\ref{Fourier})

\begin{multline}
\hat g(x) = 
\int_{\mathbb R^d}
\frac {e^{-i\mathbf{pr}}}{(2\pi)^{\frac d2}}
\frac{e^{-\frac{(\mathbf r- \mathbf r_0)^2}{2\sigma^2}}}
     {(2\pi\sigma^2)^{\frac d2}} 
\mathrm d^d\mathbf r
=
\frac 1{(2\pi)^{\frac d2}}
\int_{\mathbb R^d}
\frac{e^{-\frac{(\mathbf r + i \mathbf p \sigma^2)^2}{2\sigma^2}}}
     {(2\pi\sigma^2)^{\frac d2}} 
e^{-\frac{\mathbf p^2 \sigma^2}2}
\mathrm d^d\mathbf r
= \\
= \frac 1{(2\pi)^\frac d2} e^{-\frac{\mathbf p^2 \sigma^2}2}
\label{GaussFourier}
\end{multline}

This is similar to the normal distribution with the variance 
$\frac 1{\sigma^2}$, 
except that it is not properly normalised, 
since the Fourier transform preserves the $L^2$-norm, but not the $L^1$-norm.
For the Gaussian distribution the direct Fourier transform coincides with the inverse Fourier transform.

The Fourier transform of the convolution of $n$ 
$d$-dimensional Gaussian distributions (\ref{fenFourierGen})

\begin{equation}
\hat g_n(\mathbf p) = (2\pi)^{\frac{(n-1)d}2} 
    \frac 1{(2\pi)^{\frac {nd}2}} e^{-\frac{\mathbf p n \sigma^2}2}
= \frac 1{(2\pi)^{\frac d2}}
  e^{-\frac{\mathbf p^2 n \sigma^2}2}.
\end{equation}

\paragraph{
The pdf of the convolution of $n$ Gaussian pdfs
}
, which corresponds to the sum of $n$ normally distributed variables,
can be obtained by taking the inverse Fourier transform using (\ref{GaussFourier})
:

\begin{equation}
g_n(\mathbf r) = \frac 1{(2\pi n \sigma^2)^\frac d2}
  e^{-\frac{\mathbf r^2}{2 n \sigma^2}}.
\end{equation}

This is again the normal distribution with the variance re-scaled to $n\sigma^2$.
We use the average sum of Gaussian vectors $\mathbf r_n = \frac{\mathbf{r}}n$, 
and we shift the center of the distribution to $\mathbf r_0$;
then the standard deviation becomes $\frac \sigma {\sqrt n}$:

\begin{equation}
G_n(\mathbf r_n) = 
\frac {n^{\frac d2}}{(2\pi\sigma^2)^{\frac d2}}
e^{-\frac{n(\mathbf r_n - \mathbf r_0)^2}{2\sigma^2}}
\end{equation}

\subsection
%[CDF(cos(theta))]
{\texorpdfstring{$\mathrm{CDF}_G(\cos\theta({\mathbf r_n}))$}
{CDF(cosθ(rₙ))}}

In this subsection we work with $d = 3$. 
For calculation of integrals in spherical coordinates in arbitrary dimension one may consult \cite{Fi3}.

\begin{equation*}
\mathrm{CDF}(\cos\theta)
=
\frac{2\pi n^{\frac 32}}{(2\pi\sigma^2)^{\frac 32}}
\int_0^\infty r^2 \ud r 
\int_{\cos\theta}^1
e^{-\frac{n\left(r^2 - 2 r r_0 \cos\theta' + r_0^2\right)}{2\sigma^2}}
\ud \cos\theta'
,
\end{equation*}

the inner integral is taken easily in 3-dimensional space,

\begin{equation*}
\int_{\cos\theta}^1
e^{\frac{2 n r r_0 \cos\theta'}{2\sigma^2}}
\ud \cos\theta'
=
\frac{\sigma^2}{nrr_0}
\left(
  e^{\frac{2nrr_0}{2\sigma^2}} - e^{\frac{2nrr_0\cos\theta}{2\sigma^2}}
\right)
,
\end{equation*}

therefore

\begin{equation}
\mathrm{CDF}(\cos\theta)
=
\frac{\sqrt{n}}{\sqrt{2\pi\sigma^2}r_0}
\int_0^\infty r 
\left( e^{-\frac{n}{2\sigma^2}\left(r^2 - 2 r r_0 + r_0^2\right)}
- e^{-\frac{n}{2\sigma^2}\left(r^2 - 2 r r_0 \cos\theta + r_0^2\right)}
\right)
\ud r 
.
\label{CDF_G_eq}
\end{equation}

To calculate the first term in brackets, it is sufficient to calculate the second term and put $\cos\theta = 1$.
We complete the square in the integral

\begin{equation}
\int_0^\infty r 
e^{-\frac{n}{2\sigma^2}\left(r^2 - 2rr_0\cos\theta + r_0^2\right)}
\ud r 
=
e^{-\frac{n}{2\sigma^2}r_0^2\left(1-\cos^2\theta\right)}
\int_0^\infty r e^{-\frac{n}{2\sigma^2}\left(r-r_0\cos\theta\right)^2}
\ud r 
,
\label{G_complete_sq}
\end{equation}

then the latter integral we split into two ones with $r = (r - r_0 \cos\theta) + r_0 \cos\theta$.
The first part is a total derivative with respect to $r-r_0 \cos\theta = y$,

\begin{multline*}
\int_0^\infty r e^{-\frac{n}{2\sigma^2}(r-r_0\cos\theta)^2} \ud r =
\int_{-r_0\cos\theta}^\infty y e^{-\frac{ny^2}{2\sigma^2}} \ud y
+ r_0 \cos\theta \int_{-r_0\cos\theta}^{\infty} e^{-\frac{ny^2}{2\sigma^2}} \ud y =
\\ =
\frac{\sigma^2}n e^{-\frac{nr_0^2\cos^2\theta}{2\sigma^2}}
+ r_0 \cos\theta \int_{-r_0\cos\theta}^{\infty} e^{-\frac{ny^2}{2\sigma^2}} \ud y
.
\end{multline*}

The last term can be expressed through the 
\textit{error function} (\cite{GradRyzh}, 8.250(1)):

\begin{equation}
\mathrm{erf}(x) = \frac 2{\sqrt{\pi}} \int_0^x e^{-t^2} \ud t
\label{erf}
,
\end{equation}

\begin{align}
\int_{-r_0\cos\theta}^\infty e^{-\frac{ny^2}{2\sigma^2}} \ud y 
& =
\sqrt{\frac{2\sigma^2}n}
\int_{-\frac{\sqrt{n} r_0}{\sqrt{2\sigma^2}} \cos\theta}^\infty
e^{-t^2} \ud t
\nonumber
\\ 
& =
\sqrt{\frac{\pi\sigma^2}{2n}}
+
\sqrt{\frac{\pi\sigma^2}{2n}}
\mathrm{erf}\left(\frac{\sqrt n r_0}{\sqrt{2\sigma^2}} \cos\theta\right)
\label{G_final_erf}
\end{align}

Combining \ref{G_complete_sq} and \ref{G_final_erf} into \ref{CDF_G_eq},

\begin{multline}
\mathrm{CDF}(\cos\theta) = 
\frac{\sqrt n}{\sqrt{2\pi\sigma^2}r_0}
\Bigg( % when making line break, \left and \right won't work.
  \frac{\sigma^2}n e^{-\frac{nr_0^2}{2\sigma^2}}
  + r_0 \sqrt{\frac{\pi\sigma^2}{2n}}
    \left(1 + \mathrm{erf}\left(\frac{\sqrt n r_0}{\sqrt{2\sigma^2}}\right)
    \right)
-
\\
  - e^{-\frac n{2\sigma^2}r_0^2 \left(1-cos^2\theta \right)}
  \left(\frac{\sigma^2}n e^{-\frac{n r_0^2 \cos^2\theta}{2\sigma^2}}
  + r_0 \cos\theta \sqrt{\frac{\pi\sigma^2}{2n}}
    \left(1 + 
      \mathrm{erf}\left(\frac{\sqrt n r_0}{\sqrt{2\sigma^2}} \cos\theta \right)
    \right)
  \right)
\Bigg)
\end{multline}

The first term cancels out, and we obtain the final result:

\begin{align}
\mathrm{CDF}_G(\cos\theta({\mathbf r_n}))
= & \frac 12 \Bigg( 
  1 + \mathrm{erf}\left(\frac{\sqrt n r_0}{\sqrt 2 \sigma}\right)
\nonumber
\\
  & - e^{-\frac {nr_0^2}{2\sigma^2} \left(1-cos^2\theta \right)}
  \cos\theta
  \left(
    1 + \mathrm{erf}\left(\frac{\sqrt n r_0}{\sqrt 2 \sigma} \cos\theta \right)
  \right)
\Bigg)
\label{CDF_G_exact}
\end{align}

Note that $\mathrm{CDF_G}$ depends only on one combination of parameters 
$\sqrt n \frac{r_0}{\sigma}$.

\subsection
%[Approximations of CDF(cos(theta)) and theta(cl)]
{
Approximations of 
\texorpdfstring{$\mathrm{CDF}_G(\cos\theta)$}
{CDF(cosθ)}
and \texorpdfstring{$\theta(cl)$}{θ(cl)}
}

In this subsection we consider the behaviour of $\mathrm{CDF}_G(\cos\theta)$ 
for different values of $\cos\theta$ and parameters and the behaviour of
confidence intervals ($\theta \textrm{ or} \cos\theta$) 
for given confidence levels $\mathrm{CDF}_G(\cos\theta) = cl$.

We introduce the parameter 

\begin{equation}
a = \frac{\sqrt n r_0}{\sqrt 2 \sigma}
\label{a_G}
\end{equation}

and express the CDF as

\begin{equation}
\mathrm{CDF}(\cos\theta) = 
\frac 12 
  \left(
    1 + \mathrm{erf}(a) 
    - e^{-a^2\left(1-\cos^2\theta\right)} \cos\theta
      \left(1 + \mathrm{erf}\left(a\cos\theta\right)\right)
  \right).
\label{CDF_G_a}
\end{equation}

In what follows we work with \ref{CDF_G_a}, but keep in mind 
the expression of \ref{a_G} through the original parameters of the distribution 
$n, r_0, \sigma$.

We are interested not only in limit cases, but even more in finite statistics samples.
We group the terms according to their orders and keep lower order terms explicitly.

\subsubsection
%[ theta close to 0, n large ]
{
\texorpdfstring{$\theta$}{θ}
close to 0, 
$n$ large
}

The asymptotic representation for the error function for large argument is (\cite{GradRyzh}, 8.254)

\begin{equation}
\mathrm{erf}(z) = 1 - \frac{e^{-z^2}}{\sqrt \pi z}
\left(
  \sum_{k=0}^n (-1)^k \frac{(2k-1)!!}{\left(2z^2\right)^k} 
  + \mathit O \left(|z|^{-2n -z}\right)
\right)
\label{erf_large_arg}
\end{equation}

(where $(-1)!! = 1$).
$\mathrm{erf}(a)$ tends very rapidly to 1 as $a$ increases.  
Therefore the simplest approximation would be to substitute $\mathrm{erf}$ for $1$ for large arguments.
Thus for $a\gg1$, $a \cos\theta \gg 1$

\begin{equation}
\mathrm{CDF}(\cos\theta) = 1 - 
  e^{-a^2(1-\cos^2\theta)} \cos\theta
  + \alpha_1,
\label{CDF_G_large_param}
\end{equation}

where 

\begin{equation}
\alpha_1 = 
\frac{\mathrm{erf}(a) - 1}2 -
e^{-a^2(1-\cos^2\theta)} \cos\theta \frac{\mathrm{erf}(a\cos\theta) - 1}2
= \mathit O \left( a^{-1} e^{-a^2} \right).
\label{G_alpha_1}
\end{equation}

To express $\theta$ from \ref{CDF_G_large_param} is more difficult, 
since the exponent power $a^2\left(1-\cos^2\theta\right)$ can be arbitrary.
We fix $\mathrm{CDF}(\cos\theta) = cl$,
move the term with $\theta$ to the left side 
of \ref{CDF_G_large_param},
and take logarithm 

\begin{gather}
\ln \cos \theta - a^2 \left(1-\cos^2\theta\right) =
\ln \left(1 - cl + \alpha_1 \right)
\label{G_not_big_cl}
, \\
\frac 12 \ln\left(1-\sin^2\theta\right) - a^2 \sin^2\theta =
\ln\left(1 - cl\right) + \mathrm{ln}\left(1 + \frac{\alpha_1}{1 - cl}\right)
\label{G_expand_ln}
\end{gather}

The equation \ref{G_not_big_cl} means that 
the results for lower $cl$ will be more precise than for $cl$ very close to 1, 
namely $1 - cl$ should be much more than $\alpha_1$. 
Lower $cl$ also corresponds to smaller $\theta$.

In order to solve \ref{G_expand_ln} w.r.t. $\sin\theta$, we have to take a 
reasonable assumption $\sin^2\theta \ll 1$; we introduce

\begin{equation}
\beta_1 = \mathrm{ln}\left(1-\sin^2\theta\right) + \sin^2\theta
= \mathit O(\sin^4\theta)
\label{G_beta_1},
\end{equation}

we also rewrite the last term in \ref{G_expand_ln} as

\begin{equation}
\alpha_2 = \mathrm{ln}\left(1 + \frac{\alpha_1}{1 - cl}\right)
= \mathit O(\alpha_1)
\label{G_alpha_2}
\end{equation}

Then from \ref{G_expand_ln}, \ref{G_beta_1}, \ref{G_alpha_2}

\begin{equation}
\sin^2\theta = \frac{-\mathrm{ln}(1-cl)}{\frac 12 + a^2}
+ \frac{\beta_1}{1 + 2a^2} - \frac{\alpha_2}{\frac 12 + a^2}
\label{G_sin_theta_ord_2}
\end{equation}

The equation \ref{G_expand_ln} can be solved with a better precision 
if we take into account more terms from the ln series (see e.g. \cite{GradRyzh} 1.511). Let

\begin{equation}
\beta_2 = \mathrm{ln}\left(1-\sin^2\theta\right) + \sin^2\theta + \frac 12 \sin^4\theta
= \mathit O(\sin^6\theta)
\label{G_beta_2},
\end{equation}

then \ref{G_expand_ln} transforms to a quadratic equation on $\sin^2\theta$

\begin{gather}
\frac 12 \beta_2 - \frac 14 \sin^4\theta - \left(\frac 12 + a^2\right)\sin^2\theta
= \mathrm{ln}(1-cl) + \alpha_2,
\nonumber \\
\sin^4\theta + 2\left(1 + 2a^2\right)\sin^2\theta + 4\mathrm{ln}(1-cl) 
+ 4\alpha_2 - 2\beta_2 = 0,
\nonumber \\
\sin^2\theta = 
  -\left(1 + 2a^2\right) + 
  \sqrt{(1+2a^2)^2 - 4\mathrm{ln}(1-cl) -4\alpha_2 + 2\beta_2}
\label{G_sin_theta_ord_4}
\end{gather}

The most precise formula for big $a$ should be \ref{G_sin_theta_ord_4}. 
For very big $a$ and small $\theta$ 
we can get a simpler expression from~\ref{G_sin_theta_ord_2}:

\begin{equation}
\theta \approx 
\frac{\sqrt{-\mathrm{ln}(1-cl)}}{a} 
\stackrel{\ref{a_G}}{=}
\frac{\sqrt{-2\mathrm{ln}(1-cl)}\sigma}{\sqrt n r_0}
\label{G_theta_large_n_simple}
\end{equation}

\subsubsection
%[ theta close to pi/2 ]
{
%$\pi - \theta \ll 1$
\texorpdfstring{$\theta$}{θ}
close to 
\texorpdfstring{$\frac{\pi}2$}{π/2}
}
One can expect confidence intervals 
to be 
near $\frac{\pi}2$ when $a$ is neither too big nor too small.
Therefore in this subsubsection we assume $a \sim 1$, so that $a\cos\theta \ll 1$.

The error function for small arguments can be approximated using integration of
the exponent series in
\ref{erf} term by term:

\begin{equation}
\mathrm{erf}(z) = \frac 2{\sqrt{\pi}} \left(z - \frac{z^3}{3} + 
\mathit O\left(z^5\right) 
\right)
\label{erf_small_arg}
\end{equation}

\begin{gather}
\mathrm{CDF}(\cos\theta) = 
\frac 12 (1 + \mathrm{erf}(a)) 
-\frac{e^{-a^2}}2 \cos\theta(1 + \frac 2{\sqrt\pi} a\cos\theta + \gamma_1) 
\nonumber \\
+ \frac{e^{-a^2}}2 \left(1 - e^{a^2\cos^2\theta}\right) \cos\theta
(1 + \mathrm{erf}(a\cos\theta)), \\
\textrm{where } 
\gamma_1 = \mathrm{erf}(a\cos\theta) - \frac{2}{\sqrt\pi} a\cos\theta = \mathit O(a^3\cos^3\theta)
\label{G_gamma_1}
\end{gather}

To find $\cos\theta(cl)$ 
we denote

\begin{equation}
\delta_1 = 
(e^{a^2\cos^2\theta} - 1)(1 + \mathrm{erf}(a\cos\theta))
= \mathit O(\cos^2\theta),
\label{G_delta_1}
\end{equation}

then

\begin{gather}
\cos\theta \left(1 + \frac2{\sqrt\pi} a\cos\theta \right)
+ \cos\theta(\gamma_1 + \delta_1)
=
(1 + \mathrm{erf}(a) - 2 cl) e^{a^2} 
\label{G_small_cos_eq}
\end{gather}

In the leading order the solution $\cos\theta$ of \ref{G_small_cos_eq} 
is the r.h.s. of 
\ref{G_small_cos_eq}. 
Therefore when we solve that equation up to 
$\mathit O(\cos^3\theta)$, we chose the `+' root:

\begin{equation}
\cos\theta = \frac{-1 + 
    \sqrt{1 + \frac8{\sqrt\pi}ae^{a^2}(1 + \mathrm{erf}(a) - 2 cl)
    - \frac8{\sqrt\pi}a \cos\theta(\gamma_1 + \delta_1)
    }}
{\frac 4{\sqrt\pi}a}
\label{G_cos_theta_ord_3}
.
\end{equation}

\subsubsection
%[ theta close to pi ]
{
%$\pi - \theta \ll 1$
\texorpdfstring{$\theta$}{θ}
close to
\texorpdfstring{$\pi$}{π}
}

\begin{equation}
\mathrm{CDF}(\cos\theta) = \frac12(1 + \mathrm{erf}(a)) 
  - \frac12 e^{-a^2\sin^2\theta} \cos\theta (1 + \mathrm{erf}(a\cos\theta))
\label{}
\end{equation}

The situation when $\theta$ is close to $\pi$ can appear when we have $a$
small and we are interested in large confidence levels (our precision
is low, but still allows us to exclude a region near the pole $\theta = \pi$).
In this subsubsection we don't take assumptions on $a$,
but use a Taylor series expansion of $\mathrm{erf}(z)$ at an arbitrary point:

\begin{equation}
\mathrm{erf}(a + \Delta) = \mathrm{erf}(a) + \frac2{\sqrt\pi}e^{-a^2}\Delta
+ \mathit O(\Delta^2)
\label{erf_Taylor}
.
\end{equation}

Therefore 

\begin{gather}
\mathrm{erf}(a\cos\theta) = 
\mathrm{erf}\left(-a\sqrt{1 - \sin^2\theta}\right)
= 
- \mathrm{erf}(a) + \frac1{\sqrt\pi}ae^{-a^2}\sin^2\theta + \varepsilon_1
\label{G_varepsilon_1}
, \\
\varepsilon_1 = \mathit O(\sin^4\theta)
\nonumber
\end{gather}

To find $\theta(cl)$ we solve the equation

\begin{multline}
e^{-a^2\sin^2\theta}(-\cos\theta)(1 + \mathrm{erf}(a\cos\theta))
= 
2cl - 1 - \mathrm{erf}(a)
,\\
-a^2\sin^2\theta + \frac12\mathrm{ln}(1-\sin^2\theta) 
+ \mathrm{ln}(1+\mathrm{erf}(a\cos\theta)) = \mathrm{ln}(2cl - 1 - \mathrm{erf}(a))
\end{multline}

Using \ref{G_varepsilon_1},

\begin{align}
\mathrm{ln}(1 + \mathrm{erf}(a\cos\theta)) & =
\mathrm{ln}(1 - \mathrm{erf}(a)) 
+ \mathrm{ln}\left(1 + 
  \frac{\frac1{\sqrt\pi} ae^{-a^2}\sin^2\theta + \varepsilon_1}
  {1 - \mathrm{erf}(a)}
\right)
\nonumber
\\
& = \mathrm{ln}(1-\mathrm{erf}(a)) + \frac{ae^{-a^2}}{\sqrt\pi(1 - \mathrm{erf}(a))}
\sin^2\theta + \varepsilon_2,
\label{G_varepsilon_2}
\end{align}

\begin{equation*}
\varepsilon_2 = \mathit O(\sin^4\theta).
\end{equation*}

Using \ref{G_beta_1} and \ref{G_varepsilon_2}, we obtain

\begin{multline}
\sin^2\theta \left(
  -a^2 - \frac 12 + \frac{ae^{-a^2}}{\sqrt\pi(1 - \mathrm{erf}(a))} 
\right)
= 
-\frac{\beta_1}2 - \mathrm{ln}(1 - \mathrm{erf}(a)) - \varepsilon_2
\nonumber
\\
+ \mathrm{ln}(2cl - 1 - \mathrm{erf}(a)),
\end{multline}

\begin{equation}
\sin^2\theta = \left(
  a^2 + \frac 12 - \frac{ae^{-a^2}}{\sqrt\pi(1 - \mathrm{erf}(a))}
  \right)^{-1}
\left(\mathrm{ln}\left(\frac{1-\mathrm{erf}(a)}{2cl - 1 - \mathrm{erf}(a)}\right)
+ \frac{\beta_1}2 + \varepsilon_2 \right)
.
\label{G_theta_near_pi_ord_2}
\end{equation}

The r.h.s. of \ref{G_theta_near_pi_ord_2} is positive, since 
the argument of the last ln is larger than 1:
$1 - \mathrm{erf}(a) > 2cl - 1 - \mathrm{erf}(a)$.
However, the denominator of the logarithm's argument should also be positive,

\begin{equation}
cl > \frac{1 + \mathrm{erf}(a)}2
\label{G_theta_pi_cl}
.
\end{equation}

This means that if we want to exclude some percentage of the outcomes of 
$r_n$
with the directions near the pole, 
we should chose a confidence level which satisfies \ref{G_theta_pi_cl}.

This is a necessary, but not a sufficient condition on $cl$. 
A more precise condition is that the r.h.s. of~\ref{G_theta_near_pi_ord_2} 
is less than $1$. 

When $a$ is small we obtain
$
2\mathrm{ln}\left(\frac1{2cl-1}\right) < 1
\textrm{, and } cl > \frac12(1+e^{-1/2}) \approx 0.80.
$

%% file: direction_reconstruction.bbl
\begin{thebibliography}{99}
\bibitem{Chooz99}
Determination of neutrino incoming direction in the CHOOZ experiment 
and its application to Supernova explosion location by scintillator detectors //
M. Apollonio et al., 1999
\href{http://lanl.arxiv.org/abs/hep-ex/9906011v1}{http://lanl.arxiv.org/abs/hep-ex/9906011v1}

\bibitem{MardiaJupp}
Mardia, K. and Jupp, P.E., \textit{Directional Statistics}, Wiley, 2000. 

\bibitem{GradRyzh}
Gradshteyn, I.S. and Ryzhik, I.M., \textit{Table of Integrals, Series, and Products}, 7th ed., Academic Press, 2007.

\bibitem{GKP}
Graham, R.L., Knuth, D.E. and Patashnik, O., \textit{Concrete mathematics. A Foundation for Computer Science}, 2nd ed., Addison-Wesley, 1994.

\bibitem{PBM}
Prudnikov, A.P., Brychkov, Yu.A., and Marichev, O.I., \textit{Integrali i riady} (Integrals and series), vol. 3, \textit{Spetsialniye funktsii. Dopolnitelniye glavi} (Special functions. Additional chapters), 2nd ed., Moscow, Fizmatlit, 2003.

\bibitem{StirlingsFormula}
\href{https://en.wikipedia.org/wiki/Stirling\%27s_approximation}{https://en.wikipedia.org/wiki/Stirling's\_approximation}, retrieved 21.10.2014.

\bibitem{Fi3}
Fichtenholz, G.M., \textit{Kurs differentsialnogo i integralnogo istschisleniya} (A course on differential and integral calculus), vol. 3, 676, Moscow, Fizmatlit, 2001.

\bibitem{wErF}
\href{https://en.wikipedia.org/wiki/Error_function}{https://en.wikipedia.org/wiki/Error\_function}, retrieved 02.10.2014.

\bibitem{wIntegral_parametrique}
\href{https://fr.wikipedia.org/wiki/Int\%C3\%A9grale_param\%C3\%A9trique#.C3.89tude_locale}
{https://fr.wikipedia.org/wiki/Intégrale\_paramétrique}
, retrieved 30.10.2014.

\bibitem{CLT}
\href{https://en.wikipedia.org/wiki/Central_limit_theorem}
{https://en.wikipedia.org/wiki/Central\_limit\_theorem}
, retrieved 10.04.2015.


\end{thebibliography}
